\documentclass[graybox]{svmult}

\usepackage{mathptmx}       
\usepackage{helvet}         
\usepackage{courier}        
\usepackage{type1cm}        
\usepackage{makeidx}         
\usepackage{graphicx}        
\usepackage{multicol}        
\usepackage[bottom]{footmisc}
\usepackage{hyperref}        
\usepackage{soul}            
\hypersetup{colorlinks=true,urlcolor=blue}
\usepackage[square,numbers]{natbib}
\makeindex             

\newcommand{\nustar}{\emph{NuSTAR}}
\newcommand{\rxte}{\emph{RXTE}}
\newcommand{\hxmt}{\emph{Insight-HXMT}}
\newcommand{\nicer}{\emph{NICER}}
\newcommand{\maxis}{MAXI~J1820$+$070}
\newcommand{\gray}{$\gamma$-ray}
\newcommand{\xmm}{\emph{XMM-Newton}}
\newcommand{\suzaku}{\emph{Suzaku}}

\begin{document}
\title*{Black holes: Timing and spectral properties and evolution}
\author{Emrah Kalemci \thanks{corresponding author} and Erin Kara and John A. Tomsick}
\institute{Emrah Kalemci \at Sabanc\i\ University, Faculty of Engineering and Natural Sciences, Tuzla, Istanbul, 34956, Turkey, \email{ekalemci@sabanciuniv.edu}
\and Erin Kara \at MIT Kavli Institute for Astrophysics and Space Research, Cambridge, MA, 02139, USA, \email{ekara@mit.edu}
\and John A. Tomsick \at Space Sciences Laboratory, 7 Gauss Way, University of California, Berkeley, CA 94720-7450, USA, \email{jtomsick@berkeley.edu}}

%
%
\maketitle
\abstract{We review the timing and spectral evolution of black hole X-ray binary systems, with emphasis on the current accretion-ejection paradigm. When in outburst, stellar mass black hole binaries may become the brightest X-ray sources in the sky. Analysis of high signal to noise data has resulted in a general framework of correlated X-ray spectral and fast timing behavior during an outburst. We utilize recent data from small but powerful observatories launched in the last decade supported by multi-wavelength ground-based observations. Coordinated observations showed that outflows (in the form of jets and winds) are an integral part of this evolution, providing a coherent phenomenological picture that we discuss in terms of the hardness-intensity diagram and spectral states. We pay particular attention to the evolution of broad and narrow emission and absorption lines and hard tails in the energy spectrum, quasi-periodic oscillations, lags and reverberation from fast timing studies, making the connections with multi-wavelength observations when relevant. We use the bright outburst of \maxis\ as a recent test case to discuss different aspects of spectral and timing evolution, but the data and results are not limited to this source. In the second part of the review, we discuss competing theoretical models that can explain different aspects of the rich phenomenology. Data from future missions and simulation results will have the power to resolve discrepancies in these models and black hole binary research will continue to be an exciting field that allows for tests of fundamental physics and studies of the properties of matter in strong gravitational fields. }

\section{Keywords} 
black hole physics, accretion, accretion disks, X-rays: binaries, jets

\section{Motivation}
\label{sec:motivation}

\begin{figure*}
\includegraphics[width=0.8\textwidth,clip, bb=10 38 490 580]{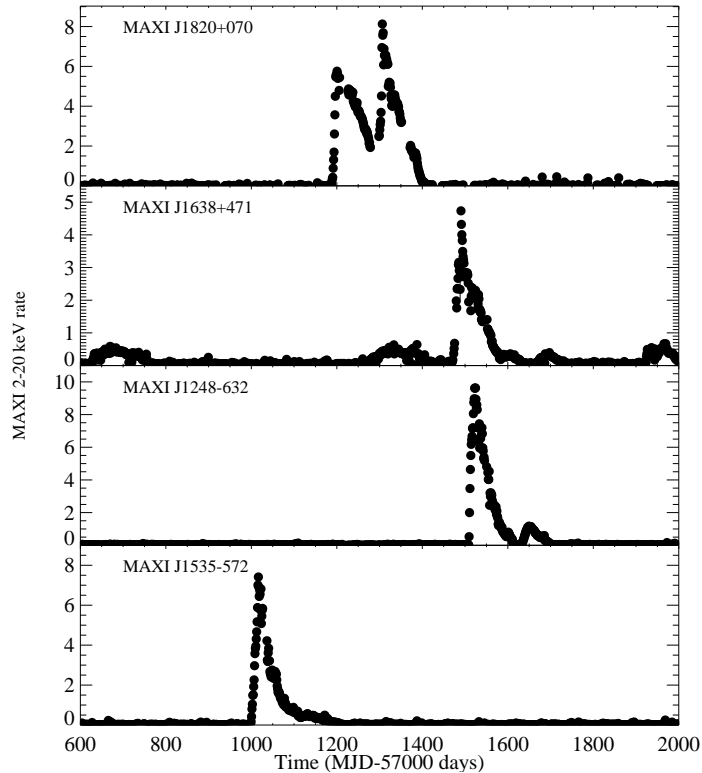}
\caption{MAXI light curves of four bright outbursts observed between MJD 57600 and MJD 59000. }
\label{fig:maxilc}
\end{figure*}

Galactic black hole (BH) research is in the midst of an era of frequent discoveries thanks to the maturation of data from small, yet powerful space observatories launched in the last decade, as well as flurry of activity from bright transients lighting up the X-ray and $\gamma$-ray sky (see Fig.~\ref{fig:maxilc}). These new observatories, namely, \emph{Nuclear Spectroscopic Telescope Array} (\nustar \cite{Harrison2013}), \emph{Neutron Star Interior Composition Explorer} (\emph{NICER} \cite{Gendreau2016}), and \emph{Hard X-ray Modulation Telescope} (\hxmt\ \cite{Zhang14}) 
provided exciting new results over those presented in the seminal review of timing and spectral evolution in X-rays by \cite{Remillard2006}, and subsequent reviews of different aspects of the subject \cite{Belloni2010, Gilfanov2010, Belloni2016, Bambi2018, Bambi2020}. 

Multi-wavelength campaigns have become the norm, with unprecedented cooperation between groups all over the world to coordinate ground- and space-based observations to probe the physics of accretion and ejection around BHs. One of these campaigns for transient source \maxis\ utilized all the active X-ray observatories at the time with multi-wavelength support. This source was not only extremely bright, but also extremely rich in terms of phenomenology observed in other sources which resulted in the publication of many interesting articles. We used \maxis\ as a case study here for the discussion of different aspects of spectral and timing evolution of BHs to highlight contributions of the new observatories on top of the established framework based on the pioneering research done in the \emph{Rossi X-ray Timing Explorer} (\rxte\ \cite{Bradt93}) era.

The review starts with an overview of spectral and timing observational results in \S\ref{sec:overview} that provides the rich phenomenology of black hole research. Next, deeper information is provided on the important aspects of the phenomenology on the observational level, leaving the discussion of the underlying models in \S\ref{sec:interpretation}. The review finishes with a discussion of what we can expect in the future with the future observatories in \S\ref{sec:conclusion}.

\section{Galactic black holes: an observational view}

\label{sec:overview}

Black hole binary (BHB) systems consist of a black hole and an ordinary star. Together with neutron star binaries they form a larger class of objects called X-ray binaries (XRB) in which matter accreted from the companion star may form an accretion disk around the compact object. The inner parts of the accretion disk heat up to millions of degrees in the deep gravitational potential well of the compact object resulting in X-ray emission. XRBs, especially the BHBs provide natural laboratories where one can, among other things, can study the effects of General Relativity in strong gravitational field through spectral and timing studies. In this review, BHB denotes systems with stellar mass black holes, intermediate mass black hole candidates are not considered. 

The first ever discovered BHB is Cyg~X$-$1, which is a persistent system and has a high mass companion (therefore it is a high mass X-ray binary). Since then, many other BHBs have been discovered. Currently, the number of BHBs reported is close to 70 (see \url{https://www.astro.puc.cl/BlackCAT/} \citep{CorralSantana2016} for an up to date list and references). Most BHBs turned out to include low mass companions (hence most are low mass X-ray binaries). Moreover, most BHBs are transients, normally their luminosities are very low ($L_{\rm x} < 10^{33}$ erg/s), and they can only be detected for a limited amount of time (typically weeks to months) when the mass accretion rate increases (see Fig.~\ref{fig:maxilc}). One can compare this luminosity with the Eddington Luminosity $L_{Edd}$, which is the maximum luminosity for which the radiation pressure balances the gravitational force towards the compact object. The Eddington Luminosity is proportional to mass of the object, and for one solar mass ($M_{\odot}$) it is 1.26 $\times 10^{38}$ erg s$^{-1}$ \citep{Bambi2018}. Note that dynamically measured masses of BHBs are typically between 5-10 $M_{\odot}$ \citep{CorralSantana2016}.

A full historical account of the discoveries regarding BHBs and their main properties is out of scope of this review, interested readers can consult \cite{Remillard2006, Belloni2016}, and references therein. The launch of \rxte\ in 1995 deserves a special mention, as it revolutionized the field thanks to the large effective area detectors with unprecedented time resolution, and an all sky monitor (ASM) that allowed many new outbursts to be detected. Its scheduling flexibility, together with improved signal-to-noise led to many discoveries and established a phenomenological landscape that is used in this review as well. In the \rxte\ era, one of the most important discoveries about the BHBs has been the overall accretion - ejection paradigm; relativistic outflows detected as jets in the radio band (\S\ref{sub:radionirjet}), and winds detected in X-rays (\S\ref{sub:winds}) were found to be closely related to X-ray spectral and temporal behavior in an outburst.  

The fast timing properties of BHBs are studied in the Fourier domain through the power spectral density (PSD), frequency-dependent lags and coherence techniques (\S\ref{sub:qpo}, \S\ref{sub:lags}). The PSD show both broad and narrow features, and the narrow features are called quasi-periodic oscillations (QPOs). 

\begin{figure*}
\includegraphics[width=\textwidth]{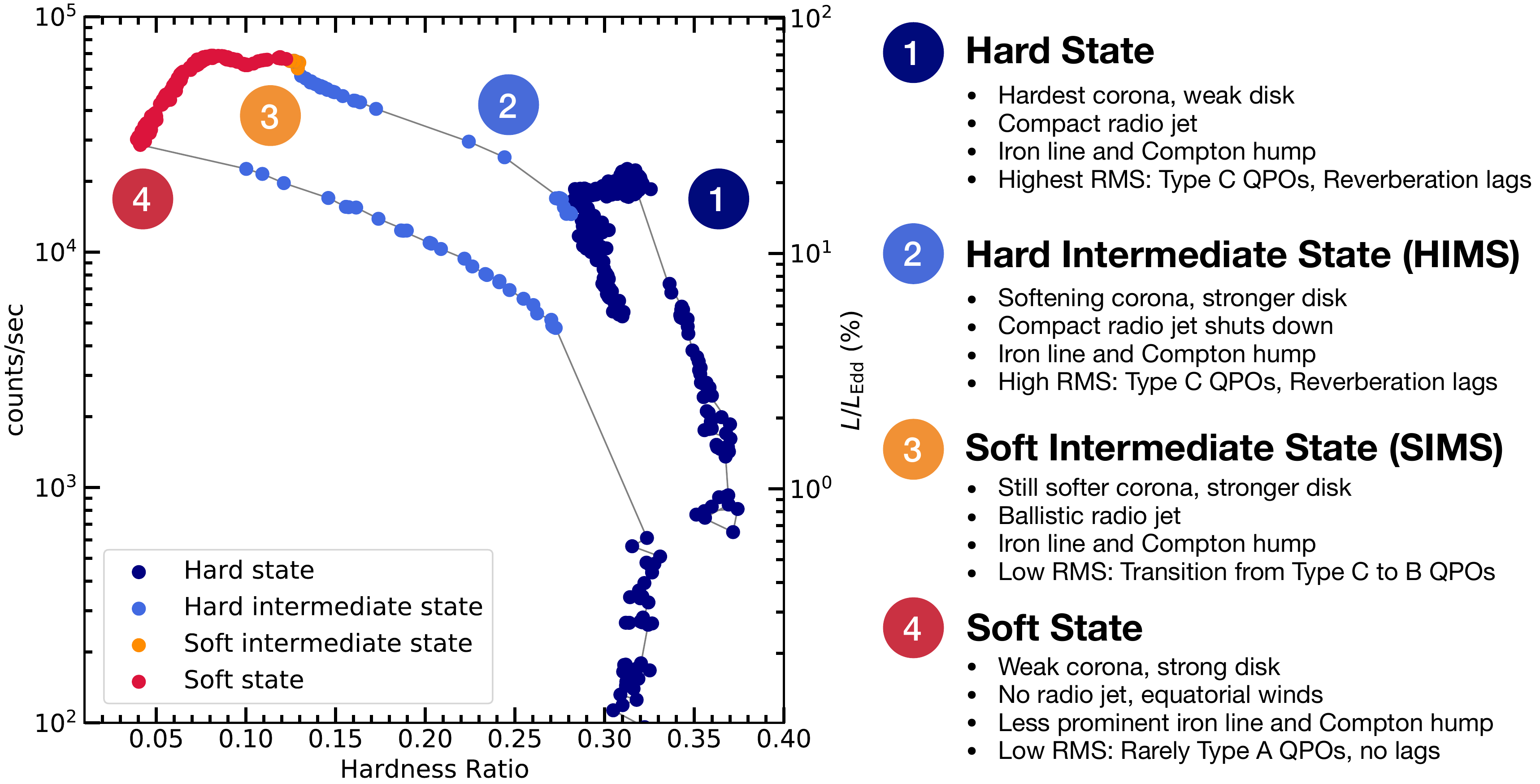}
\caption{The hardness intensity diagram of \maxis, illustrating the four main accretion states of a low mass X-ray binary outburst. For each state, we provide a short description of the key observables. Figure credit: Jingyi Wang and authors.}
\label{fig:maxi_hid}
\end{figure*}

Most of the time, BHBs are in a very low mass accretion regime called the \emph{quiescence}, or the {\emph{quiescenct state}}. But in an outburst, they may span eight orders of magnitude in luminosity, with typical quiescence levels of below $10^{-5} L_{\rm Edd}$ \citep{Plotkin2013} to maximum levels of a few $L_{\rm Edd}$ \citep{Motta2017}. This places them amongst the brightest X-ray emitters in the sky. During outbursts, BHBs often follow a well known track when the intensity is plotted against the X-ray hardness called the ``Hardness Intensity Diagram'' (HID) \cite{Belloni2016} as shown in Fig.~\ref{fig:maxi_hid}. In this track, sources show distinct and correlated X-ray spectral and fast timing properties called states. Moreover, the behavior in other wavelengths (especially in radio and near infrared) is also closely related to the X-ray spectral states.

\begin{figure*}
\includegraphics[width=\textwidth]{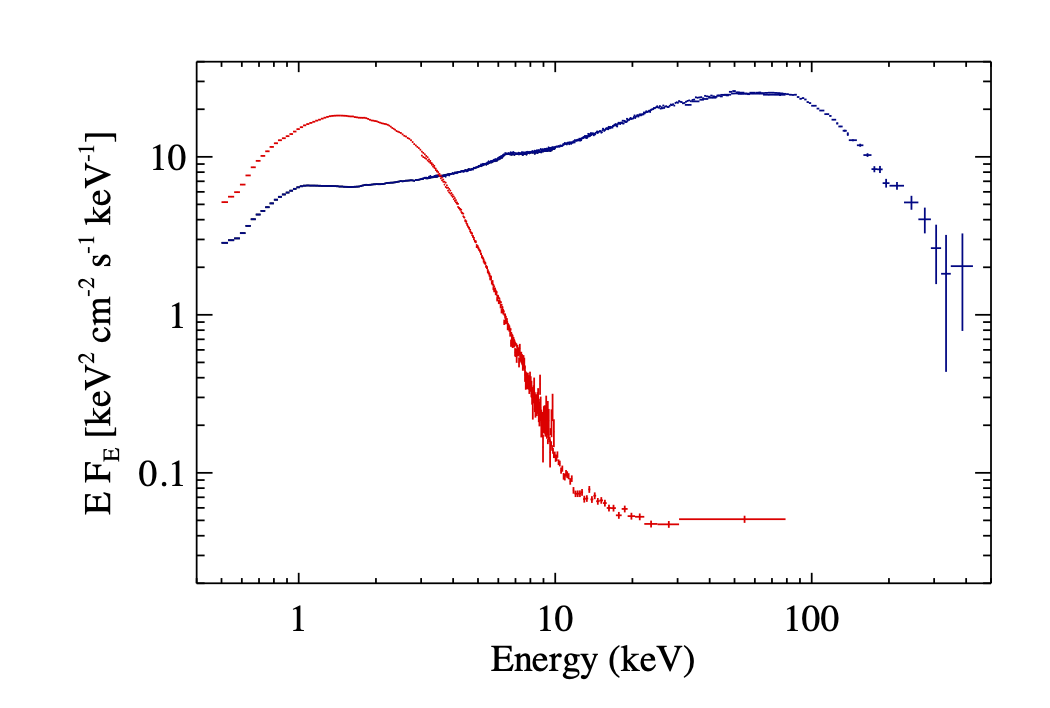}
\caption{The X-ray and $\gamma$-ray spectra of \maxis\ at two times during its 2018 outburst shown in log-log scale. The red and blue points show the spectrum of the source in the soft and hard states, respectively.}
\label{fig:maxi_spe}
\end{figure*}

Galactic BHs tend to travel counterclockwise in the HID, from the bottom right, completing a sequence of states before fading into quiescence (see \S\ref{sub:outliers} for a discussion of sources not following the usual sequence). The outbursts start in the {\emph{hard state}} in which the X-ray spectrum (the navy blue curve in Fig.~\ref{fig:maxi_spe}) is dominated by a hard component associated with inverse Comptonization of seed photons from inner parts of the disk in a hot electron corona (see \S\ref{sub:coronajet}; \S\ref{sub:geometry} for a discussion of the origin and geometry) with a distinct break at tens to hundreds of keV (\S\ref{sub:gammaray}). While a softer component associated with blackbody emission from the disk may exist (\S\ref{sub:diskmodel}), its contribution to the X-ray luminosity is low. 

A component due to the reflection of hard photons by the disk is often present, and part of this component is an emission line due to fluorescence of iron.  The shape of the line provides a means to study the effects of strong gravity on the inner accretion disk material.  A broad and often asymmetric iron line from reflection of disk material close to the black hole is often seen at higher luminosities but can turn into a narrow iron line at low luminosities (\S\ref{sub:broadironlines}).

The power spectral density (PSD) reveals strong broad-band noise (tens of \%\ RMS) and quasi-periodic oscillations (QPOs, see \S\ref{sub:qpo} for details). The observations in the hard state for \maxis\ are shown in navy blue color in the HID (Fig.~\ref{fig:maxi_hid}). The radio spectrum in this state is consistent with compact, self absorbing jets \citep{Fender2001} (\S\ref{sub:radionirjet}).

As the outburst evolves with increasing accretion rate, the soft emission associated with the disk increases, and a transition to the {\emph{hard intermediate state}} (HIMS) is observed. The emission from the corona becomes softer as well (if modelled with a power-law spectrum, the spectral index increases compared to those in the hard state). Type~C QPOs are still present and the overall root mean square variability (RMS) is still high, but the RMS drops with increasing disk flux (see Fig.~\ref{fig:maxi_psd}). 

In some sources, there is a sudden transition to a softer state called the {\emph{soft intermediate state}} (SIMS) manifested with a radio flare, a decreased continuum RMS amplitude of variability and appearance of a Type~B QPO (see Fig.~\ref{fig:maxi_psd}, \S\ref{sub:radionirjet}, and \S\ref{sub:qpo}). Some sources show rapid transition back and forth between the HIMS and the SIMS \cite{Belloni2011}. The observations in HIMS and SIMS for \maxis\ are shown in blue and orange, respectively, in the HID (Fig.~\ref{fig:maxi_hid})

As the disk emission becomes more dominant, another transition takes place to the {\emph{soft state}}. In this state, the emission from the corona is very weak (the contrast can be seen in Fig.~\ref{fig:maxi_spe}). The soft \gray\ spectrum extends without a break up to a few MeVs \cite{Grove1998}. The RMS variability is very low (a few \%\ RMS, see Fig.~\ref{fig:maxi_psd}) and QPOs are absent. The observations in the soft state for \maxis\ are shown in red in the HID (Fig.~\ref{fig:maxi_hid}). Radio emission is quenched by 3.5 orders of magnitude \citep{Russell2019}, indicating that the jets are absent; however, high resolution X-ray spectral observations indicate presence of an equatorial wind in high inclination sources \citep{Ponti2012} (\S\ref{sub:winds}). 

As the disk flux starts to decrease, the source may go through the SIMS and HIMS and eventually ends up at the hard state again \cite{Belloni2016}. The transition back to the hard state takes place at a much lower flux compared to the transition to the softer states during the outburst rise (sometimes this behavior is called "hysteresis"). There are also indications that the transition back to the hard state occurs at a fixed bolometric luminosity \citep{Maccarone2003, Vahdat2019}. As the luminosity goes even lower, some sources show a softening in their spectrum before they fade to quiescence \citep{Kalemci2013, Plotkin2013}.  

A few important additional items should be noted at this point. For papers before and in the early \rxte\ era, the hard state and the soft state were denoted as the low state and the high state, respectively. While using "high/low" instead of "soft/hard" state could be justified if only the soft X-rays are considered, when X-ray bolometric luminosities are used, the hard state rise luminosities can exceed the soft state luminosities \cite{Dunn2010}. This is also evident in Fig.~\ref{fig:maxi_spe}. Sometimes terms like high soft state and low hard state are used as well \cite{Belloni2016}. A few sources divert to a very luminous (close to or exceeding the Eddington Luminosity) and soft state called the ultra-luminous, hyper-soft or anomalous state \cite{Belloni2010}. For in depth discussion of different definitions of states, see \cite{Belloni2010}.

While the hardness intensity diagram is extremely useful to see the big picture in terms of spectral states, it does not provide the details that may be very important to characterize the source behavior. For example, it is not clear how fast the evolution takes place in different states; while the intermediate states span a wide range in hardness in \maxis\ (see Fig.~\ref{fig:maxi_hid}), they lasted for 20 days. In contrast, the source spent more than 100 days in the initial hard state \cite{Shidatsu2019}. For other useful diagrams utilized in describing especially the timing properties of states, see references in \cite{Belloni2016}.

\subsection{Iron lines}
\label{sub:broadironlines}

The X-ray spectra of accreting BHs often show a reflection component that is formed when hard X-rays irradiate the disk.  This component includes a fluorescent iron emission line, an iron absorption edge, and a broad excess peaking at tens of keV \citep{lw88,fabian89}.  Relativistic velocities of the disk material and a redshift from the BH's gravitational field distort the reflection component \citep{laor91}.  Although the the entire reflection component is affected, the distortion is most apparent in the iron K$\alpha$ emission line, and this is clearly seen in both Active Galactic Nuclei \citep{tanaka95} and stellar mass black holes \citep{miller07}.

Although sources do not show reflection components at all times, and some sources do not show them at all, the presence of reflection is common.  The majority of sources have reflection components, and they can occur in a variety of spectral states.  For example, Cyg~X$-$1 exhibits a strong reflection component in its hard state \citep{nowak11,parker15}, its soft state \citep{walton16}, and its intermediate state \citep{tomsick18}.  GX~339$-$4 is an active BH transient with a well-studied reflection component in multiple spectral states \citep[e.g.,][]{zdziarski98,parker16}.  

\begin{figure*}
\includegraphics[width=1.0\textwidth]{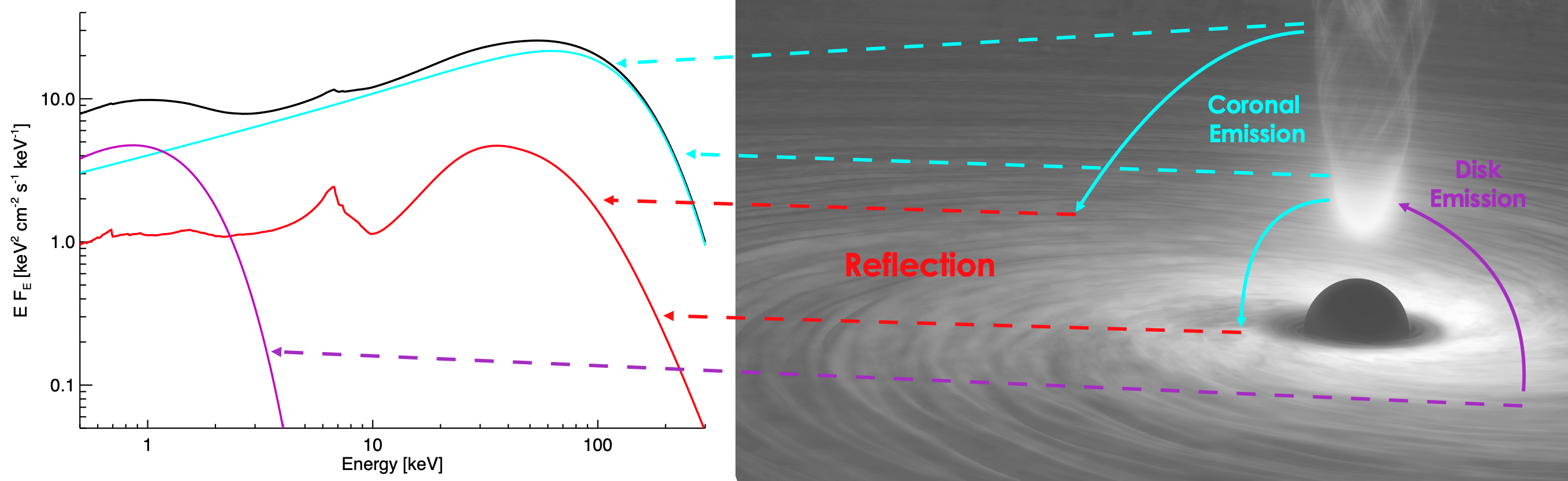}
\caption{A possible physical scenario for the X-ray production based on the model used to fit hard state spectra of \maxis\ in \cite{Buisson19}.  We thank Dr. Javier Garcia for providing a first version of this figure.}
\label{fig:cartoon}
\end{figure*} 

The velocities in the inner portions of the disk can exceed 10\% of the speed of light, producing a symmetric distortion of close to 1\,keV at the 6.4\,keV energy of the iron emission line.  The gravitational effect produces an asymmetric distortion, which leads to a red wing of the emission line.  A main goal of modeling the iron line and reflection component is to measure the location of the inner edge of the accretion disk ($R_{\rm in}$). If the disk extends to the innermost stable circular orbit (the ISCO), then measuring $R_{\rm in}$ provides a measurement of the BH spin.  Even if the disk is slightly truncated and does not reach the ISCO, the $R_{\rm in}$ measurement still provides a lower limit on the BH spin.  

Some spectral and timing measurements can be interpreted as the disk becoming highly truncated for sources in the hard state \citep{tomsick09,plant15,demarco16,xu20}, while other measurements are not indicative of truncation \citep{miller15}.  Most observations that suggest a high level of truncation have occurred for sources that are at low luminosities near 0.1\% Eddington.  At these low luminosities, the iron line profile changes from having a relativistic profile to a relatively narrow line, which provides a low limit on $R_{\rm in}$ \citep{tomsick09,plant15,xu20}.  However, the relativistic iron line seen at high luminosities in the hard state leaves it unclear precisely when the disk becomes truncated.  Determining the location of $R_{\rm in}$ and characterizing its evolution in the hard state is an active area of research \citep[e.g.,][]{wang18}.  

The iron line and other spectral components are illustrated in Fig.~\ref{fig:cartoon}.  The spectrum shown of \maxis\ in the hard state is based on modeling of a \nicer\ and \nustar\ spectrum.  The two direct components are the thermal disk emission at low energies and the coronal emission, which is modeled as being produced by Comptonization by high-energy electrons in the jet with a thermal distribution.  Fig.~\ref{fig:cartoon} illustrates the source geometry used in work by \cite{Buisson19}.  We discuss possible source geometries, which are likely state-dependent, in \S\ref{sub:geometry}.  The properties of the reflection component provide motivation for the double-lamppost since many sources have reflection components that combine relativistic reflection from the inner disk and non-relativistic reflection from the outer disk.  

\nustar\ has enabled advances in reflection and iron line studies for a number of reasons.  \nustar's 3--79\,keV bandpass is very well-matched to measuring the reflection component.  Its advantages over \rxte\ are its much better energy resolution as well as improved sensitivity (especially at the high-energy part of the bandpass) due to lower background. \nustar\ also provides advances over CCD instruments like those on \xmm\ and \suzaku\ by providing sensitive measurements at higher energies.  In addition, photon pile-up can lead to spectral distortion in CCD instruments \citep{done10,miller10} which is not trivial to account for.  Although pile-up can often be mitigated by using certain detector modes or by modeling the pile-up effects, it is a significant source of systematic uncertainty for bright sources.  However, \nustar\ has a triggered readout system that is not susceptible to pile-up.  The \nustar\ advances in bandpass and avoiding pile-up have occurred along with advances in reflection modeling by making the models more physically realistic \citep{garcia14}.  These advances have increased the reliability of the modeling technique as a tool to measure the properties in the inner accretion disk, including the inner disk radius and the spins of black holes.

\subsection{Absorption lines and winds}
\label{sub:winds}

\begin{figure*}
\includegraphics[width=0.45\textwidth]{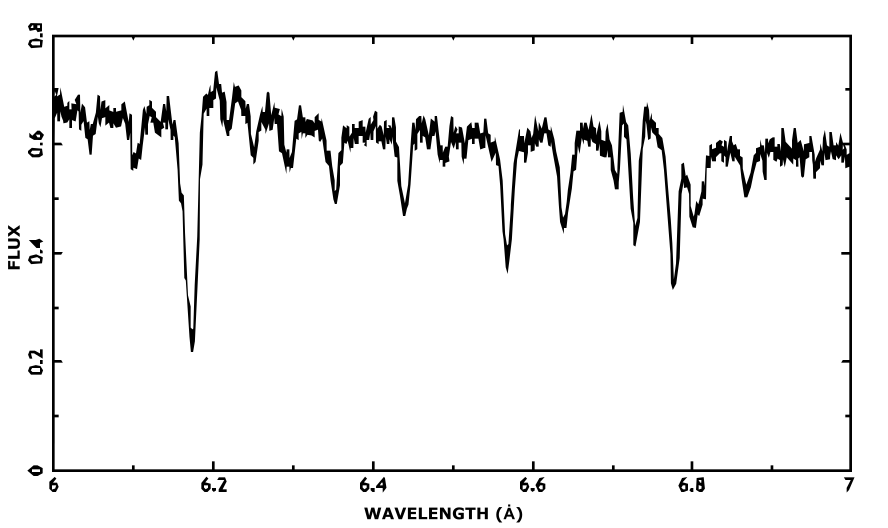}
\includegraphics[width=0.55\textwidth]{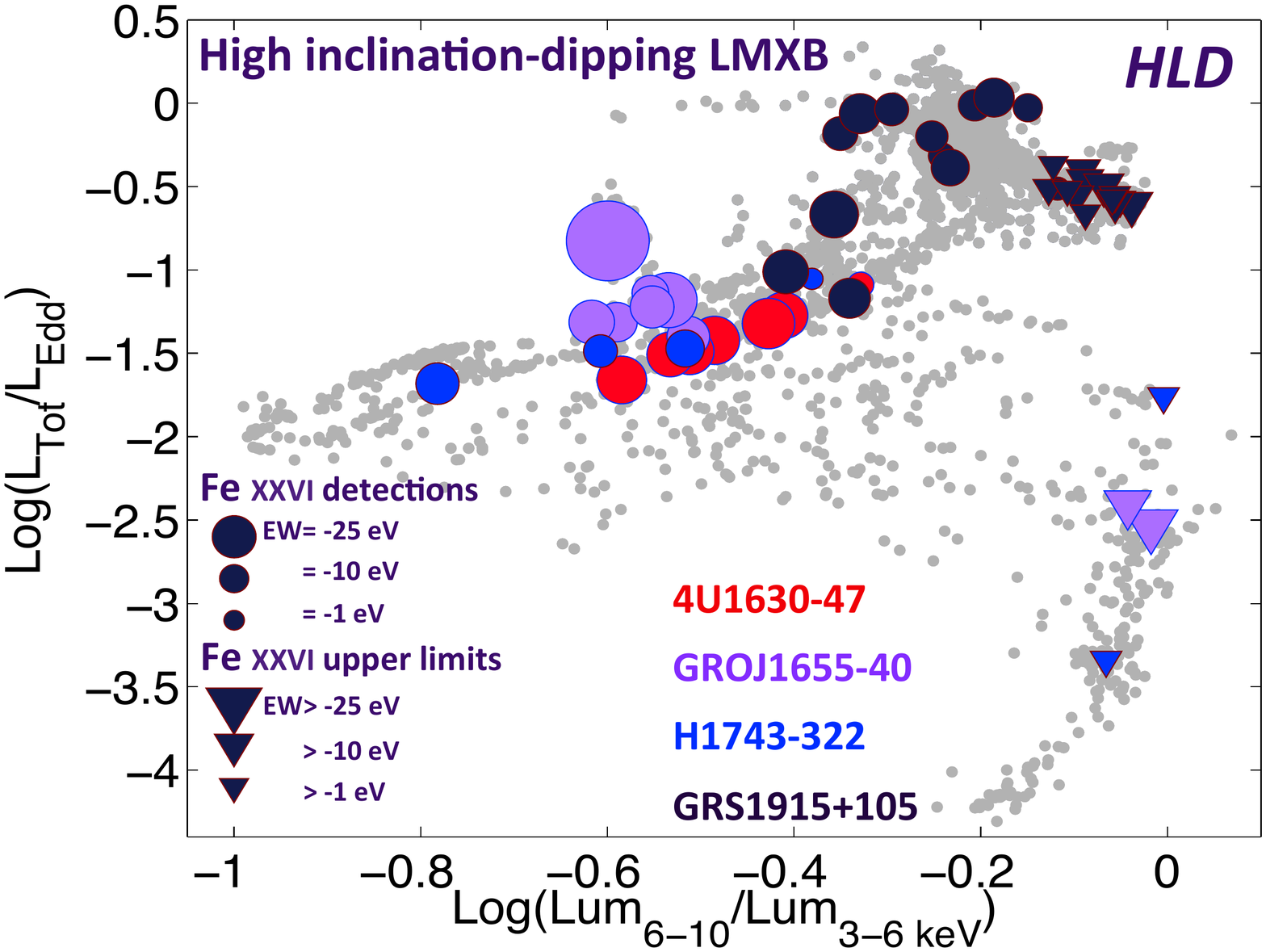}
\caption{Left: Chandra grating spectrum of GRO~J1655$-$40 in the soft state. Figure credit: Adapted, NASA/CXC/U.Michigan/J.Miller et al. \cite{Miller2006}. Right: The hardness luminosity diagram of high inclination LMXBs, combined with detection (circles) and upper limits (triangles) of key wind absorption feature. Gray circles show all observations from all sources, and colored circles and triangles are from high inclination BHBs. Figure credit: Ponti et al. \cite{Ponti2012}.}
\label{fig:ponti}
\end{figure*}

High resolution spectral observations of BHBs revealed the presence of absorption lines (see Fig.~\ref{fig:ponti}, Left) blue shifted by up to a few thousand km\,s$^{-1}$, showing that these sources not only produce collimated jets but can also drive winds. The lines are mainly hydrogen- and helium-like iron K${\alpha}$, indicating material that is strongly photoionized by the X-ray illumination from the central source (see reviews by \cite{Ponti2012, Ponti2016, DiazTrigo2016}).

The wind is an important component in the accretion flow as the mass outflow rate in this component can exceed the mass accretion rate through the disk \citep{Ponti2012, King2012}. 
It is also one of the determining factors in the general outburst profile and duration \citep{Dubus2019, Tetarenko2020}. Winds are  observed across the black hole mass scale \citep[][and references therein]{King2013}, and a comparison of their properties in supermassive BHs and Galactic BHs can provide clues about the driving mechanisms of the winds.

Whether we can observe evidence of winds in X-rays depends on the spectral state and disk inclination: wind signatures are preferentially detected in soft states for high inclination sources \citep[see Fig.~\ref{fig:ponti},][]{Done2007, Ponti2012}. The inclination effect indicates that the plasma has a flat, equatorial geometry above the disk providing large amount of material in the line of sight. Wind signatures are also imprinted in the optical emission lines of He and H as P Cygni profiles and broad emission line wings \citep[][and references therein]{MunozDarias2019}. These ``cold winds" are detected in the hard state. In fact, similar near infrared line features obtained by VLT/X-Shooter indicate that the cold winds are present throughout the outburst in both soft and hard states \citep{Sierras2020}. See \S\ref{sub:windorigin} for a discussion of competing models to explain the origin of winds in BHBs.

\subsection{Radio and near infrared emission and jets}
\label{sub:radionirjet}

In addition to the accretion disk, corona and winds observed in the X-ray band, XRBs are also known to launch powerful jets that are most clearly observed through synchrotron emission in the radio to infrared (IR). These jets are capable of carrying away a significant amount of the accretion energy and depositing large amounts of energy into their surrounding environments \cite{gallo05,russell07,tetarenko17}. Radio emission is only detected in the hard state and during the state transition.  Thus, it is clear that the launching mechanism is somehow related to the accretion process, but the causal connection between accretion and ejection and how these relativistic jets are launched, collimated and accelerated remain open questions. Here we describe the properties of the radio and IR emission during a the black hole outburst, and observational efforts to probe the disk-jet connection.

At the beginning of an outburst, when the X-ray emission is spectrally hard, the radio and IR show a steady, compact (AU-sized) jet \cite{dhawan00,corbel00,Fender2001,stirling01,Fender2004}. The radio-to-mm spectral energy distribution (SED), shows a flat or inverted spectrum that extends up to $\sim 10^{13}$~Hz, beyond which the jet becomes optically thin, and the spectrum steepens. This frequency of the spectral break is thought to be set by the location in the jet where particle acceleration begins: higher frequencies correspond to an acceleration zone that is closer to the black hole. In a typical outburst, as the accretion rate increases throughout the hard state, the spectral break is observed to evolve to lower frequencies (down to the radio band), which has been interpreted as being due to the particle-acceleration region moving further from the black hole.

During the transition from hard to soft state, the steady, compact jet turns off, but that is not the end of the story for the jet. In several cases where high-cadence radio monitoring was possible, the end of the state transition was marked by rapid flaring in radio before completely shutting down in the soft state. The flares have been associated with the ejection of discrete knots of plasma moving away from the black hole at relativistic velocities (e.g. \cite{mirabel94, fender04, bright20}). No compact radio core is observed in the soft state, but one can observe residual and spatially extended radio emission from ejecta launched during the state transition. In some cases, these knots can reach separations of tens of thousands of times farther than the jet core. In the remarkable case of \maxis, \cite{bright20} discovered that during the state transition, an isolated radio flare from the compact core expanded to a spatially extended, bi-polar, relativistic outflow that was also detected in the X-rays, likely due to shocks from the jet ejecta interacting with the interstellar medium \citep{corbel02,espinasse20}. 

The dramatic change in both the jet (as observed in radio) and the inner accretion flow (as probed in X-rays) during an outburst motivated studies correlating these two wavelengths in order to better understand their connection. Indeed, the strong correlation between X-ray and radio luminosity suggests that an increase in the mass accretion rate onto the black hole results in an increase in mass loading in the jet \cite{hannikainen98,corbel02,gallo03}. This same correlation is even seen in supermassive black holes when scaled properly for black hole mass \cite{merloni03,falcke04}. While early studies suggested X-ray binaries in outburst exist on a single track, such that $L_{\mathrm{radio}} \propto L_{\mathrm{X}}^{\sim 0.7}$ \cite{corbel02}, in a large population study, \cite{gallo12} showed that there were two populations: one that is ``radio-loud" and another ``radio-quiet", which may be related to changes in the radiative efficiency of the accretion flow throughout the outburst \cite{casella09,koljonen18,espinasse18} and/or variations in observed luminosity due to inclination affects \cite{zdziarski16,motta18,munosdarias13,heil15,Motta2015}. 

Beyond correlations in population studies, cross-correlation between the X-ray band and radio/IR/optical bands on timescales as short as milliseconds, have revealed insights into the causal connection between these emitting regions. Several systems show a rapid IR/optical lag of $\sim 0.1$~s behind the X-rays \cite{gandhi08,gandhi17,vincentelli19,paice19}, which has been interpreted as being due to the propagation delay between the X-ray corona and the first IR/optical emission zone in the jet, and thus constrains the physical scale over which plasma is accelerated and collimated in the inner jet \cite{jamil10,malzac13,malzac14}. This model can also explain the observed lag on a timescale of $\sim$minutes between radio and sub-mm bands, as the propagation front moves outwards in the jet. In addition to these lags, other tantalizing broadband timing properties are observed in several sources, but their origins remain a topic of debate, including: an anti-correlation of the X-ray and optical/IR at zero-lag (e.g., \cite{veledina11,paice21}) and quasi periodic oscillations (QPOs) at same temporal frequency in both X-ray and optical/IR \cite{vincentelli21}. Simultaneous, fast photometry observations in X-rays and longer wavelengths are challenging to coordinate, and the field is relatively young. Future coordinated campaigns at several points in the outburst have the potential to be a powerful tool for probing disk-jet connection in X-ray binaries. 

\subsection{Quasi-periodic oscillations}
\label{sub:qpo}

\begin{figure*}
\includegraphics[width=\textwidth]{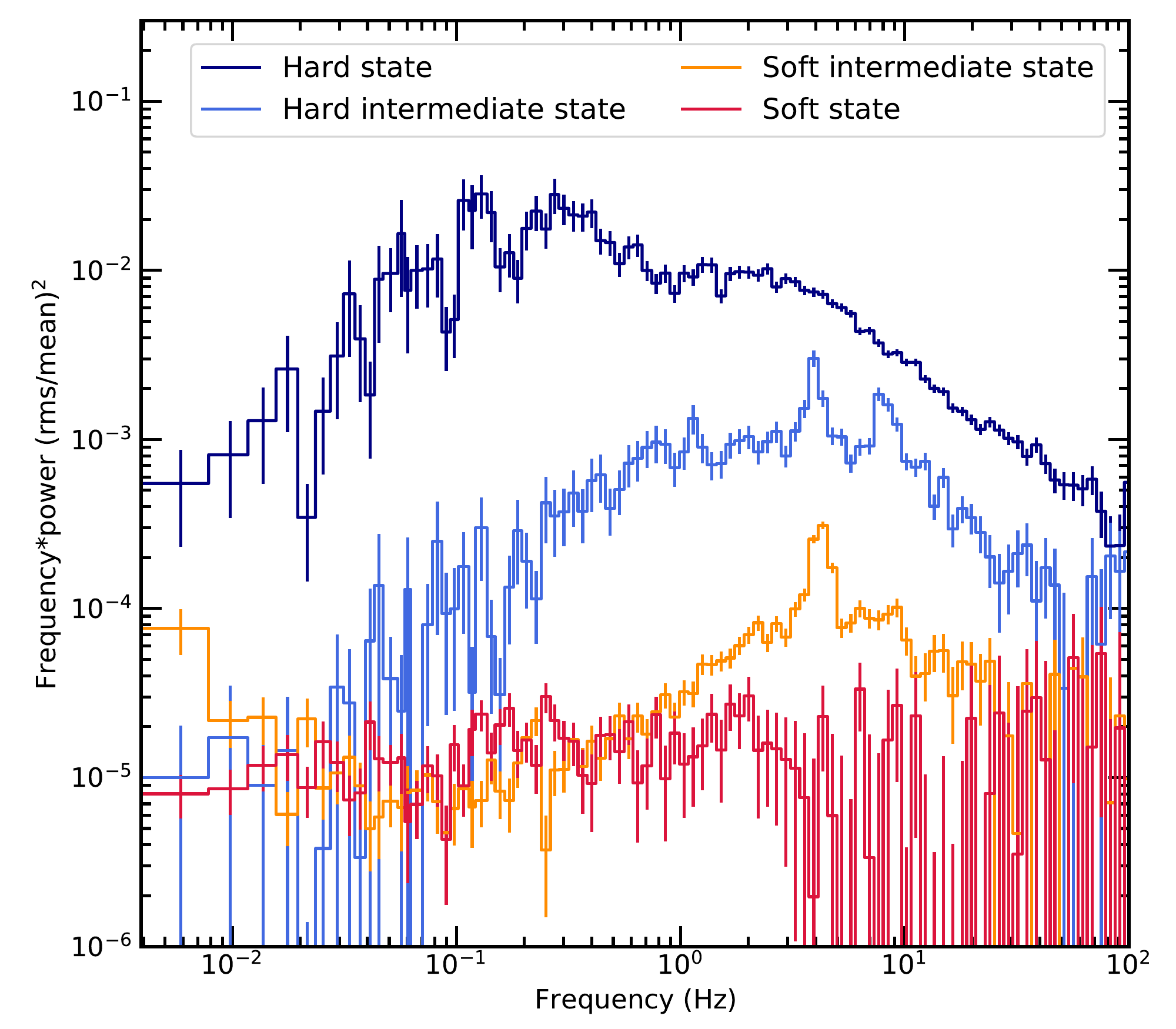}
\caption{The power spectral density of \maxis\ in four spectral states. The colors match the spectral types in Fig.~\ref{fig:maxi_hid}. Figure credit: Jingyi Wang.}
\label{fig:maxi_psd}
\end{figure*}

QPOs are a tantalizing phenomenology commonly observed during black hole outbursts, where variability power is concentrated in a narrow frequency range (Fig.~\ref{fig:maxi_psd}). Their origin is still debated, but much progress has been made in the last decade to understand the physics behind these striking features. QPOs broadly take the form of ``low-frequency QPOs" (LFQPO, in the temporal frequency range of $\sim 0.005-40$~Hz) and ``high frequency QPOs" (HFQPO, observed at $\sim 40-450$~Hz). There is a rich phenomenology and an admittedly confusing nomenclature, which we attempt to distill below. We refer readers to more thorough descriptions of QPOs in \cite{casella05}, \cite{Belloni2010} and \cite{Ingram2019}.

\subsubsection{Low-Frequency QPOs}

As the source emerges in the hard state (dominated by the hard Comptonization component), the variability is very strong over a wide range of timescales (resulting in a broadband noise PSD that is often phenomenologically modelled as several broad Lorentzian components with fractional RMS of $\sim 30$\%). Just as the spectra are observed to change dramatically, so too do the variability properties. As the source flux increases in the hard state, the variability becomes more rapid, resulting in a shifting of the broad Lorentzians to higher frequencies. Often the variability becomes more concentrated at particular frequencies, resulting in the appearance of coherent QPOs. These particular hard and HIMS state QPOs are known as ``Type~C QPOs," and are by far the most common type of QPO observed in black hole XRBs. They occur in a frequency range of $\sim 0.05-30$~Hz, often accompanied by their higher order harmonics. 

Type~C QPOs have been known for decades (first seen with {\it Ariel~6} observations of GX~339$-$4; \cite{motch83}), and their origin remains elusive, but a major break-through came in 2015 when \cite{Motta2015} and \cite{heil15} discovered that the strength of the Type~C QPO is dependent on the inclination of the binary orbit. More edge-on systems show stronger Type~C QPO amplitudes. This suggests that the origin of Type~C QPOs is geometric rather than due to intrinsic resonances in the accretion flow. Moreover, \cite{ingram16} found in one source, H1743-322, that the iron line centroid energy shifts with QPO phase, which gives stronger credence to geometric models. Possible origins of Type~C QPOs are discussed in detail in Section~\ref{sub:qpoorigin}.

Entering the HIMS, more power is concentrated in the QPOs and the broadband noise level diminishes to a fractional RMS value of $\sim 10$\%. The frequency of the Type~C QPOs continues to increase throughout the HIMS, until suddenly the Type~C QPO vanishes and is replaced by a Type~B QPO, which typically occurs in a narrower frequency range (1--6~Hz) and generally has a lower variability amplitude than Type~C QPOs. The presence of the Type~B QPO marks the transition to the SIMS. Type~B QPOs are less commonly observed than Type~C QPOs, but this may simply be due to the fact that they are short-lived. In \maxis\, the Type~C QPO (and harmonics) were observed for $\sim 100$~days, while the Type~B QPO was seen in half of one observation for just a few hundred seconds. The Type~B QPO has long been suspected to be associated with the transient ballistic jet seen in the SIMS, and in \maxis\, \cite{Homan2020} showed the clearest connection to date between these phenomena. The Type~B QPO was detected 1.5--2.5~hours before the radio jet flare. Type~B QPOs also show an inclination dependence (again suggesting a geometric origin), but unlike Type~C QPOs, they are stronger in face-on systems. 

The soft state, when the accretion disk dominates the spectrum, shows very little variability on the $\sim$hour timescale of a typical observation (i.e. a fractional RMS of $<3$\%). The PSD is often described by a single broad Lorentzian or broken powerlaw, usually with no low-frequency QPOs, though in the RXTE era, a handful of so-called Type~A QPOs were found \citep{Wijnands99}. No Type~A QPOs have been seen by NICER, and \citep{casella05} suggest Type~A QPOs are, in fact, a subset of Type~B QPOs. As the source transitions back to the hard state, at the end of the outburst, the Type~C QPOs return, and the broadband variability increases again. 

\subsubsection{High-Frequency QPOs}

HFQPOs have properties that are distinct from LFQPOs.  In addition to having higher frequencies, they tend to appear near certain characteristic values, such as the 67\,Hz HFQPO seen from GRS~1915$+$105 \citep{Morgan97} and the 66\,Hz signal from IGR~J17091$-$3624 \citep{Altamirano12}.  They are relatively weak with typical RMS amplitudes of several percent or less; however, their strength often increases strongly with energy \citep{Morgan97,Remillard2006,Strohmayer01}.  They have been detected at high significance by \rxte\ from at least four accreting BH systems (the two mentioned above, GRO~J1655$-$40, and XTE~J1550$-$564), and at lower significance for five or six more \citep{Remillard2006,Belloni12}.  They have also been detected by {\em AstroSat} for GRS~1915$+$105 \citep{sreehari20}.

For GRO~J1655$-$40, by searching the 13--30\,keV band, \cite{Strohmayer01} found a signal at 450\,Hz, which is the highest frequency that has been observed.  When combined with the measured mass of the BH in this system, the frequency corresponds to an orbital time scale at a radius smaller than the ISCO of a non-rotating BH, providing evidence that the GRO~J1655$-$40 BH must have non-zero spin \citep{Strohmayer01,Motta14}.  Thus, these signals are of high interest as they probe the innermost regions of the accretion disk.  The origin of these signals is not known, but theoretical models have been developed to explain why systems with multiple signals have frequencies that cluster around a 3:2 ratio, such as the 41 and 67\,Hz HFQPOs in GRS~1915$+$105, the 300 and 450\,Hz HFQPOs in GRO~J1655$-$40, or the 180 and 280\,Hz HFQPOs in XTE~J1550$-$564.  However, while the 3:2 ratio was originally thought to be a defining characteristic of HFQPOs, further analysis has shown that at least some (and perhaps most) of the 3:2 ratios are spurious \citep{Belloni12,Ingram2019}.

HFQPOs are relatively rare, and, in most cases, they were found only by intense monitoring of sources with \rxte.  \cite{Belloni12} find that the XTE~J1550$-$564 HFQPOs are most often seen when the source is in the SIMS, but they conclude that HFQPOs are not only seen in the SIMS.  However, the detection of HFQPOs does appear to require that the source has strong thermal and power-law components \citep{Remillard2006}.  \rxte's combination of hard X-ray bandpass and large effective area enabled it to obtain HFQPO detections.  {\em NICER}'s soft X-ray bandpass is not as well-suited, and it is also difficult for \nustar\ to detect HFQPOs due to its smaller effective area than \rxte's Proportional Counter Array and its relatively high level of deadtime for bright sources. \hxmt, with its broad energy band, large detection area and favorable pile-up properties is well suited for HFQPO detection \citep{Zhang14}, however, no HFQPO detection from BHBs has been reported with this observatory at the time of writing this review.

\subsection{Lags and reverberation}
\label{sub:lags}

\begin{figure*}
\includegraphics[width=0.43\textwidth]{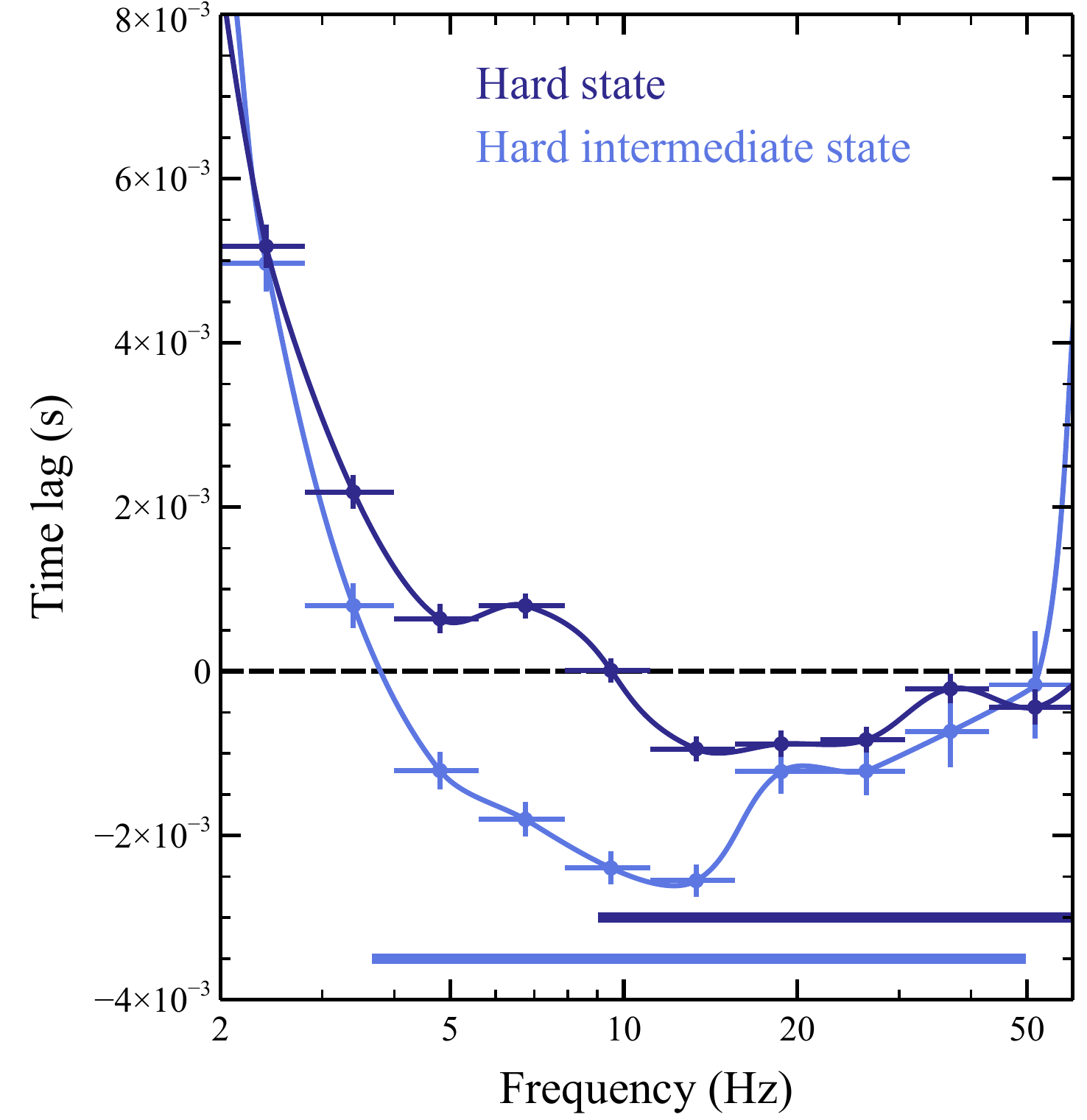}
\includegraphics[width=0.55\textwidth]{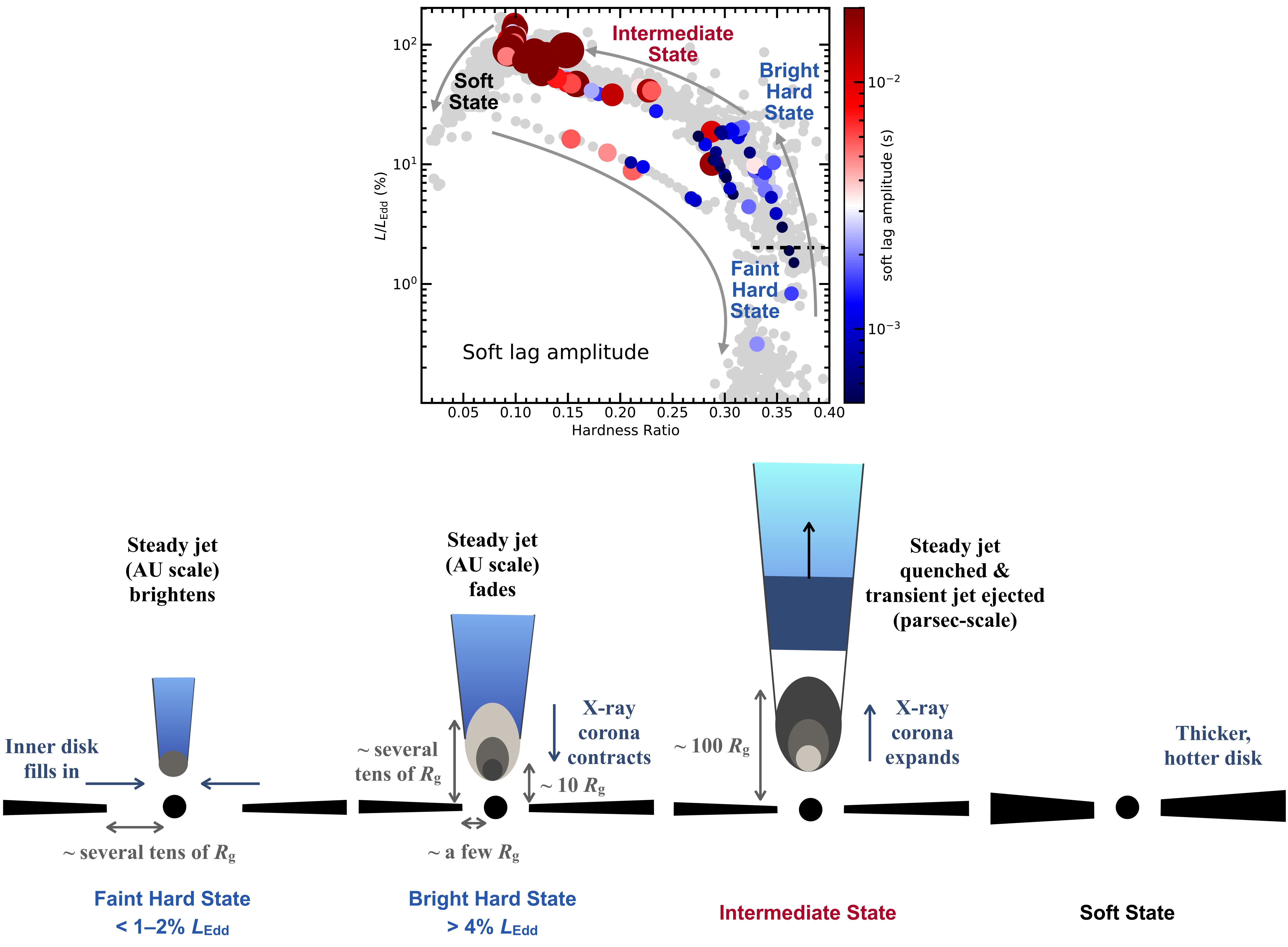}
\caption{Left: The frequency-dependent time lags between the `soft band' (0.3--1~keV) and the `hard band' (1--4~keV) for \maxis\ in one hard state observation and the HIMS. By convention, the soft lag is defined to be negative, and is the feature interpreted as due to reverberation lags between the coronal continuum and reflection off the accretion disk. The colors match the spectral types in Fig.~\ref{fig:maxi_hid}. The soft lags become larger in amplitude and lower in frequency during the state transition. Figure credit: Jingyi Wang. Right: The results of a systematic search for soft reverberation lags in 10 black hole LMXBs (\cite{wang22}), where the HID for all sources has been linearly scaled to that of \maxis\ . The colored markers indicate sources where soft lags were detected, and the color and marker size indicate the amplitude of the lag. The lags are shortest in the bright hard state, and become significantly longer during the state transition.}
\label{fig:maxi_lag}
\end{figure*}

While the power spectrum is the amplitude of the Fourier transform of a given light curve, much information is also contained in the phase, and in particular the phase difference between different energy bands (see Fig.~\ref{fig:maxi_lag}). For a detailed review on Fourier-resolved time lag techniques and modeling, we encourage readers to see \cite{uttley14}. Here, we provide a short history of the field, with particular focus on new results from XRBs and how the lags evolve with spectral state.

Fourier-resolved time lags in stellar-mass black holes were first discovered with \rxte\ in the hard state of the high-mass X-ray binary, Cyg~X$-$1 \cite{miyamoto88,nowak99,pottschmidt00}. In this source, hard photons were observed to lag soft photons on long timescales (low Fourier-frequencies $<1~Hz$). Indeed, similar {\em low-frequency hard lags} have also been seen in Active Galactic Nuclei (first in \cite{papadakis01}), and during the hard state in low-mass X-ray binaries \cite{uttley11}, but are not typically observed in the soft state of LMXBs, when variability amplitudes are too low. The low-frequency hard lags are typically 0.1~s or longer, and thus are unlikely to arise from a light travel time lag, which would imply that the corona is thousands of gravitational radii ($R_{\rm g}$ = $GM/c^{2}$)  \cite{nowak99}. Thus the prevailing interpretation is that these lags are related to the radial inflow (i.e. viscous) timescale \cite{kotov01}, as mass accretion rates propagate inwards through the accretion disk \cite{lyubarskii97}. As such, these lags are often referred to as {\em propagation lags} or {\em continuum lags}.

More recently, time lag studies have expanded to higher Fourier frequencies, revealing a new phenomenology: {\em the high-frequency soft lag.} First discovered with \xmm\ in the hard state of GX~339$-$4 \cite{uttley11}, these lags are interpreted as due to the light travel time between the X-ray corona and the accretion disk, though how the observed lags translate to the exact truncation radius of the disk or the height of the corona is still debated (see Section~\ref{sub:geometry}). \cite{demarco15} expanded these studies of GX~339$-$4 in the rise in the hard state, and found that the lag amplitude decreased during the hard state, which was interpreted either as the truncation radius of the disk moving inwards or the corona contracting before the state transition. Pile-up in \xmm\ timing mode precluded the spectral confirmation of this picture by simultaneous mapping of the iron line profile.

Now {\em NICER} has revolutionized spectral-timing studies thanks to its high throughput, good energy resolution with virtually no pile-up that allows us to measure both the time lags and spectra simultaneously. \cite{kara19} studied the luminous hard state of the bright XRB \maxis\ and discovered that before the state transition, the reverberation lags evolved to higher frequencies with a smaller amplitude lag. During this same period, the broad iron line profile remained constant, suggesting that the truncation radius did not change, and rather, the vertical extent of the corona decreased. Recently, \cite{wang21} and \cite{demarco21} discovered that during the state transition, days before the detection of a radio flare and a Type~B QPO, the reverberation lag suddenly increased in amplitude (and decreased in Fourier frequency; Fig.~\ref{fig:maxi_lag}-left). Both papers suggest this result indicates that the hard-to-soft transition is marked by the corona expanding vertically and launching a jet knot that propagates along the jet stream at relativistic velocities. Most recently, \cite{wang22} performed an automated NICER search for time lags in black hole LMXBs, and discovered that the longer reverberation lags during the state transition is not only observed in MAXI~J1820+070, but is common behaviour in black hole transients (Fig.~\ref{fig:maxi_lag}-right).

Not only are time lags associated with the broadband noise, but also distinctive lags are observed in Type~C \cite{altamirano15,Karpouzas21_GRS1915,mendez22,zhang22_J1535} and Type~B QPOs \cite{stevens18,garcia21_typeB}. Some sources show hard QPO lags, while others show soft lags, which may be determined simply by inclination effects (low orbital inclination sources show hard lags, while high-inclination sources show soft lags; \cite{vandeneijnden17}) or perhaps is indicative of time-dependent radiative properties of the corona \cite{mendez22}, where the balance of heating and cooling of the corona can lead to hard or soft QPO lags \cite{Karpouzas21_GRS1915,zhang22_J1535}. The latter model was recently applied to $Insight-HXMT$ observations of the low-frequency QPOs of MAXI~J1535$-$571 during the transition from HIMS to SIMS, and inferred that the corona extends vertically during the state transition (similar to inferences from the broadband reverberation lags described above; \cite{wang21,demarco21,wang22}). \cite{mendez22} recently compared the inference from this variable Comptonization model to the radio jet luminosity in GRS~1915$+$105, which also suggests a connection between the powering of the corona and the relativistic jet. 

High-throughput instruments with moderate energy resolution are revolutionizing our understanding of LMXBs, allowing for modeling of both the timing and spectral products. Time lags are an example of one such technique where the energy and time-dependence of the photons are accounted for in the models. Frequency-resolved spectroscopy is another technique that allows obtaining the energy spectrum of the X-ray flux variations on different time scales. It has been applied to very bright sources to probe the geometry of the black hole vicinity \citep{revnivtsev99, baby22}, as well as to study the origin of QPOs \citep{Axelsson16}. Other examples of important spectral-timing techniques include: phase-resolved spectroscopy (e.g. \cite{stevens18,ingram16}), cross-spectral analysis (e.g. \cite{mastroserio19}), and the bispectrum (e.g. \citep{arur22}).

\subsection{Soft \gray s and Polarization}
\label{sub:gammaray}

The observations with the OSSE instrument on the \emph{Compton Gamma-ray Observatory} provided the first evidence of the distinct hard X-ray (from $\sim$10 keV to a few hundred keVs) - soft \gray\ (from a few hundred keVs to a few MeVs) behaviour in the hard and the soft states of BHBs. In Cyg~X$-$1, the hard state shows a break at around 100 keV while the soft state emission has no break up to the MeV range \cite{Grove1998}. This behavior has been seen in many sources, and can also be observed for \maxis\ in Fig.~\ref{fig:maxi_spe}. The spectra are consistent with thermal Comptonization in the hard state, thereby providing a measure of the electron temperature and the optical thickness in the corona, and non-thermal Comptonization in the soft state from electrons with an initial power-law energy distribution \cite{Coppi1999}.

The game changer in understanding the high energy emission behavior of BHBs has been the $INTEGRAL$ Observatory with a combination of sensitive instruments operating in the hard X-ray and soft \gray\ bands (as well as a softer X-ray instrument) and its ability to perform uninterrupted observations with a large field of view. A recent review by \cite{Motta2021} discusses the contribution of the $INTEGRAL$ observatory in BHB research in detail, and only a few highlights are presented here. 

One of the most important contributions of $INTEGRAL$ has been the confirmation and detailed investigations of a soft \gray\ tail in the spectra of several BHBs in the hard and the hard intermediate states extending up to the MeV energies. Equally important discovery made by $INTEGRAL$ is the polarization measurement of Cyg~X$-$1 in the soft \gray\ band showing that the polarisation fraction increases with energy (from less than 20\% below $\sim$230 keV, to greater than 75\% above $\sim$400 keV, \cite{Laurent2011, Jourdain2012}). Soft X-ray (at 2.6 keV and 5.2 keV) polarization measurements of Cyg~X$-$1 performed with $OSO-8$ indicate a relatively low (compared to soft \gray\ measurements) polarization level of 2-5\% with low significance \cite{Long1980}, consistent with the decreased polarization fraction with decreasing energy.

Historical soft \gray\ band observations also claimed narrow and broad features associated with positron annihilation; a broad feature in Cyg~X$-$1 measured with $HEAO 1$ \cite{Nolan1983}, a narrow line for Nova Muscae \cite{Sunyaev1992} and a broad and variable feature in 1E1740.7--2942 \cite{Bouchet1991} measured with SIGMA on $GRANAT$. Recently, using the SPI instrument on $INTEGRAL$, a detection of a broadened 511 keV annihilation line during a strong flaring activity in the 2015 outburst of V404~Cyg has been claimed \cite{Siegert2016}, yet other groups failed to detect the same feature in the same dataset \cite{Roques2019, Jourdain2017}. Comprehensive searches for annihilation line features with IBIS and SPI on $INTEGRAL$ did not find any point source, and the emission in the associated band is consistent with diffuse positron-annihilation \cite{Decesare2011, Tsygankov2010}. \maxis\ is no exception as a detailed analysis of the soft \gray\ region extending up to 2 MeV with PICSiT and SPI on $INTEGRAL$ only provided a strong upper limit to an annihilation line feature. The non-detection, together with the Comptonization model parameters allowed a calculation of the electron-positron pair density which is turned out to be low \cite{Zdziarski2021b}.

Finally, observational campaigns have been performed during the state transitions with $RXTE$ as well as with $INTEGRAL$ in the hard X-ray band to characterize the evolution of the break frequency (or cut-off frequency depending on the spectral model used) and other spectral parameters to understand the possible physical changes in the accretion environment during transitions. GX~339$-$4 in the 2004 outburst showed a rapid transition ($<$10 hrs) from the HIMS to SIMS with significant changes happening in the break frequency and the timing properties \cite{Belloni2006}. The evolution of the break energy during the transitions are quite complex, decreasing during the hard state to the HIMS, and then increasing, disappearing, or moving beyond the detection range towards the transition to SIMS \cite{Motta2021}. 

\subsection{Outliers in hardness intensity diagram evolution}
\label{sub:outliers}

\begin{figure*}
\includegraphics[width=\textwidth]{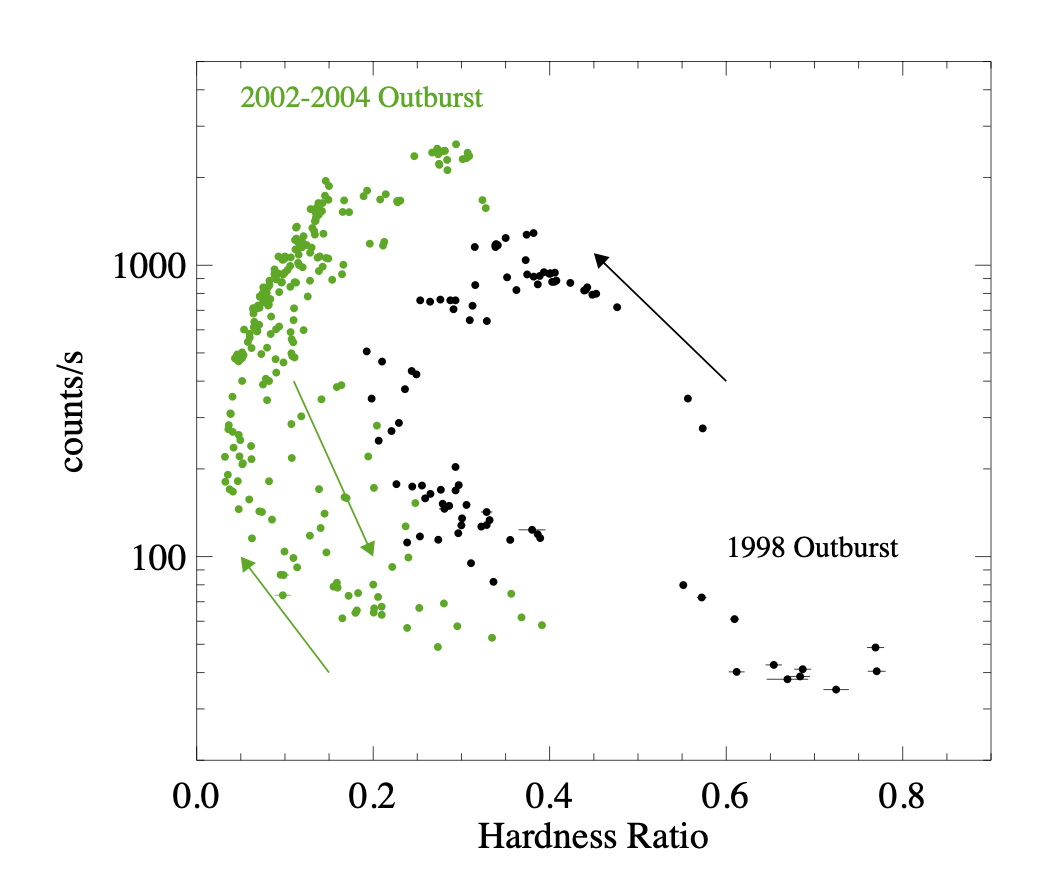}
\caption{HIDs from the BH transient 4U~1630--47 during a regular outburst from 1998 and during a clockwise outburst in 2002-2003.  The data are from \rxte, and the hardness ratio is the 9-20 keV count rate divided by the 3-9 keV count rate.  The count rate on the vertical axis is the 3-20 keV count rate.  Adapted from \citep{tomsick06}.}
\label{fig:hi_1630}
\end{figure*}

There are a significant number of exceptions to the standard path taken by BH transients in the HID.  For example, there are many cases of ``failed outbursts'' that do not reach the SIMS and soft state, and there are also ``hard-only'' outbursts that do not even make it into the HIMS \citep{Motta2021}.  These differences in the HIDs occur on an outburst-by-outburst basis,  and Fig.~\ref{fig:hi_1630} shows different outburst types for the same source.  While some sources may more often have one type or another type of outburst, a single BH transient can exhibit regular, failed, and hard-only outbursts.  Thus, the difference may be related to different mass accretion rates between outbursts or perhaps different properties of the disc (e.g., magnetic field) or jet.  

There are also some interesting accreting BH sources that have unusual aspects of their HID behavior.  Cyg~X$-$1 is a persistent source and makes transitions between the hard and soft states, but its luminosity remains unchanged to within about 15\% \citep{zhang97}.  Thus, it occupies a relatively narrow range of intensity in the HID.  GRS~1915$+$105 can show extremely rapid flux oscillations with distinctive and repeating patterns.  At times, these oscillations occur while GRS~1915+105 is restricted to the bright and soft part of the HID.  Combining X-ray, radio, and near-IR measurements provides evidence for a connection between the X-ray oscillations, which are presumably related to disk instabilities, and jet production \citep{fender04}.

Although less common, there are also cases where the hardness-intensity path differs from regular outbursts.  Low-soft states have been seen in the persistent BH systems 1E~1740.7--2942 and GRS~1758--258.  Both of these sources spend the vast majority of their time in the hard state, but they will occasionally {\em decrease} in flux and transition to the soft state \citep{smith01}.  This has also been seen in the BH transient Swift~J1753.5--0127, which showed a very similar behavior at one point during an outburst that lasted over a decade \citep{shaw16}.  It has been suggested that this phenomenon may be due to the presence of two accretion flows, a thin disk and a hot halo \citep{smith02}.  The low-soft states can be naturally produced if the time scale for changes in mass accretion rate to propagate through the disk are different in the two flows.

Another BH transient that shows low-soft states is 4U~1630--47 \citep{tomsick14_1630}.  Approximately 20 outbursts have been seen from this source going back to 1969.  In 1998, it showed a regular outburst rising in the hard state and moving in a counter-clockwise direction through the HID (Fig.~\ref{fig:hi_1630}).  However, most of the 4U~1630--47 outbursts show the opposite pattern with the source taking a clockwise path through the HID.  Although the reason for this difference is not known, it has some relationship to the the strength of the jet because 4U~1630--47 typically shows weak or undetectable radio emission during outbursts, but the strongest radio emission seen from 4U~1630--47 occurred during the 1998 outburst \citep{hjellming99}.

\section{Modelling and Interpretation}
\label{sec:interpretation}

In this section, the emphasis will be on the physical explanations and the models put forward to explain different aspects of observations discussed earlier. Due to page limitations, only major models have been discussed.  However, references to recent reviews for more comprehensive coverage are provided in each section.

\subsection{Thermal disc modelling}
\label{sub:diskmodel}

Accreting black holes often have a strong thermal component in their spectra (see Fig.~\ref{fig:maxi_spe}), especially (but not exclusively) when they are emitting at high luminosities.  This component is very well-described by a multi-temperature disk, which is approximated by adding the contributions of blackbody emission from many annuli in the disk.  A theoretical temperature vs. radius profile \citep{ss73} can be used to determine the temperature for each annulus. The simplest and often used version of this model is the ``disk-blackbody'' model \citep{mitsuda84}, and it is specified by the inner disk temperature and a normalization.  The normalization depends on the apparent inner radius, the distance to the source, and the inclination of the disk.  With corrections described in \cite{kubota98}, a physical value for the inner radius ($R_{\rm in}$) can be estimated.  Measuring $R_{\rm in}$ is of great interest since it is a key disk geometry parameter, having implications for what happens during state transitions as well as providing an opportunity to measure the spin of the BH, $a_{*}$ \citep{middleton16}.

Using the measurement of $R_{\rm in}$ to determine $a_{*}$ is only justified if the inner edge of the optically thick disk is at the innermost stable circular orbit (ISCO). In addition to the properties of the thermal component, there are other observational lines of evidence that the disc reaches near to or to the ISCO.  The red wing of the iron emission line (see \S\ref{sub:broadironlines}) shows that the emission reflected from the disc comes from the direct vicinity of the BH.  Similarly, the HFQPOs must originate from very close to the ISCO.  Measurements of $R_{\rm in}$ obtained by modeling the thermal component (``continuum fitting'') often show constant $R_{\rm in}$ values over large ranges of luminosity, strongly suggesting that the inner edge reaches a limiting value \citep{steiner10}, which is naturally interpreted as the ISCO.  

Using continuum fitting to determine $a_{*}$ requires a significant amount of additional knowledge of the binary BH system.  To determine $R_{\rm in}$ in physical units requires knowledge of the distance to the system as well as the disk inclination, which is usually assumed to be the same as the binary inclination.  Once $R_{\rm in}$ is determined in physical units, then the mass of the BH is needed to calculate $R_{\rm g}$, and determine $R_{\rm in}$/$R_{\rm g}$, which determines $a_{*}$.  Measurements of $a_{*}$ using the continuum fitting technique was accomplished for ten sources \citep{mcclintock14} with a range from $a_{*} = 0.12\pm 0.19$ (A~0620--00) to $>$0.95 (Cyg~X$-$1).  Recently, an update of the Cyg~X$-$1 distance \citep{mj21} has led to a revision of the spin estimate to $a_{*} > 0.9985$ \citep{zhao21}.  

While improved distances and masses for more systems will lead to more opportunities for $a_{*}$ measurements from continuum fitting, the systematic uncertainties need to be considered carefully.  For example, spectral hardening of the disc continuum by the disc atmosphere leads to a higher observed temperature than the effective temperature of the disc.  A correction using the spectral hardening factor, $f_{\rm col}$, is typically made when using the continuum fitting technique, but it has been argued that the uncertainty in $f_{\rm col}$ is large enough to lead to significant uncertainty in $a_{*}$ \citep{sm21}.  In addition, there have been challenges to the assumption of spin-orbit alignment \citep{walton16,sp20}. One avenue to further improvements is the joint measurements of $a_{*}$ with the continuum fitting and reflection (iron line modeling) techniques \citep{parker16}. Spectro-polarimetry is another type of measurement that is expected to provide advances in this area (see Chapter 5).

\subsection{Origin of winds} 
\label{sub:windorigin}

The two main unresolved discussion points regarding the winds in BHBs are the driving mechanism of the wind (thermal, magnetic or radiation pressure) and the disappearance of X-ray absorption lines in the hard state.

The bright soft state provides ample, possibly unobscured ionizing radiation to lift matter from the disk surface through Compton heating \citep{Begelman1983}. For this thermal driving mechanism the wind is launched at large distances (10$^4$ - 10$^{5}$ $R_{g}$) from the central black hole. The launch radius will be a fraction of the Compton radius \citep[the distance such that escape speed equals the isothermal sound speed at the inverse Compton temperature,][]{Woods1996}, and the exact value of the fraction depends on properties of the illuminating X-rays \cite{Done2018}. 

The other possible mechanisms to drive winds are radiation pressure and magnetic processes. For Galactic BHs, line driving is not efficient due to low UV flux, and high ionization \citep{Done2007}; however, under certain conditions radiation pressure can transfer momentum to free electrons \citep{Reynolds2012, Done2018}. On the other hand, magnetic wind launching mechanisms are viable possibilities. In fact, for GRO~J1655$-$40, a single observation revealed a rich series of absorption lines (see Fig.~\ref{fig:ponti}) from a dense, highly ionized wind, initially interpreted as magnetically driven \citep{Miller2006} as the calculated wind launch radius was much smaller than the Compton radius inferred for the given source luminosity prohibiting thermal winds. Magnetically driven winds close to the black hole are implied in other sources such as IGR J17091$-$3624 \citep{King2012, Miller2012} and 4U~1630$-$47 \citep{King2014}.  However, the magnetic origin in the single GRO~J1655$-$40 observation has been challenged by multiple authors. If the wind is optically thick, the luminosity will be underestimated (it could be close to Eddington) leading to complications in the interpretation of the data due to radiative mechanisms (see \citep{Done2018} for additional discussion regarding arguments against magnetic origin). Perhaps both origins are possible within a magneto-thermal hybrid wind, with magnetic launching from the inner parts of the disk and thermal launching from the outer parts of the disk in the same source \citep{Neilsen2012, Trueba2019}.

The disappearance of the X-ray absorption features in the hard state has been linked to the over-ionization of the material, photoionization instabilities, geometrical obscuration of the outer disk by a large scale-height corona, and the details of the driving mechanism \cite{Ponti2012, Chakravorty2013, Chakravorty2016}. Over-ionization in a static absorber is ruled out by detailed photoionization computations \citep{Neilsen2012}, and observationally the hard state luminosity during the rise of the outburst is similar to the soft state luminosities with wind features (see Fig.~\ref{fig:ponti}).   

Perhaps the lack of winds in the hard state and the origin of the driving mechanism of the wind are interlinked. The hard states show ubiquitous jets, and an analysis of accretion ejection behavior in GRS~1915$+$105 led to the to claim that both jets and winds are part of the same magnetic system and the degree of collimation determines whether we see a jet or wind \citep[e.g. ``jet-wind dichotomy''][]{Neilsen2009}. However, winds and optically thin jets are known to coexist \citep{Kalemci2016, Homan2016}, and cold winds may be present in all states. Some magnetohydrodynamic simulations indicate that that the inclusion of magnetic fields into a thermally driven wind has the effect of suppressing the wind in low-plasma beta regions \cite{Waters2018}. There are rare claims of hard state wind features in the X-ray spectra such as for Swift~J1658.2-–4242 with \nustar\ \citep{xu2018} and for V404~Cyg and GRS~1915$+$105 with {\em Chandra} \citep{Homan2016}. Perhaps more sensitive instruments (such as {\em Xrism} and later {\em Athena}) will provide more insight to the properties and the origin of the hard state winds. 

\subsection{Hard state accretion geometry}
\label{sub:geometry}

While the general picture for the soft state is a non-truncated optically thick disc extending to the ISCO, there is no consensus on the accretion geometry in the hard state.  In the context of the advection-dominated accretion flow (ADAF) model, a paradigm emerged where the disc begins to truncate in the intermediate state, $R_{\rm in}$ increases further in the hard state, and then reaches $10^{2}$ to $10^{4}$ Schwarzschild radii below Eddington ratios of $\sim$0.01 \citep{esin97}.  Some measurements, such as the drop in the temperature of the thermal disc component and the drop in the QPO frequencies as the luminosity drops support this picture \citep{dincer14}.  In addition, other measurements can be explained by the truncated disc picture, including the size and evolution of the X-ray lags \citep{demarco16} as well as changes in the iron line, which has been found to change from a relativistic profile to a narrow line at Eddington ratios of 0.001-0.003 \citep{tomsick09,xu20}. An analysis of the frequency-resolved spectra of \maxis\ also indicated a reflector moving closer to the inner edge of the accretion disk in the hard-to-soft transition \citep{Axelsson21}. However, the relativistic iron lines that are seen in the bright ($\sim$0.01-0.1 $L_{\rm Edd}$) hard states \citep{miller15,garcia15} strongly contradict the idea that disc truncation is necessary for sources to show hard state properties.

Another critical addition to our understanding of BHB spectral states in the last two decades is that a compact jet is present in the hard state.  The jet dominates the spectrum at radio wavelengths \citep{Fender2001} and can dominate up to at least the near-IR \citep{Corbel2002b,gandhi11}. Although it very likely extends to X-rays and perhaps gamma-rays, it is unclear what fraction of the X-ray and gamma-ray emission we see comes from the jet (see also \S\ref{sub:gammarayorigin}).  The question of how much X-ray emission comes from the jet compared to a more extended corona is important since the reflection modeling that is used to derive the inner radius of the disc depends on the source geometry \citep{dauser13}.  

As described in \S\ref{sub:broadironlines}, narrow and broad iron line components are often seen in hard state spectra, and one explanation that has been advanced for \maxis\ is that there is X-ray emission coming from different heights in the jet that illuminate both the inner and outer parts of a non- or mildly-truncated disc \citep{kara19,Buisson19}.  However, it has been shown that the \maxis\ can also be described by other geometries.  The narrow and broad iron line components have also been modeled with a moderately truncated disk and illumination by multiple X-ray components \citep{zdziarski21}.  Within the reflection interpretation, the broad iron line does not allow for a highly truncated disc, but these two analyses show that complex models can lead to different numerical answers for $R_{\rm in}$.  

In addition to the questions of disc truncation, high-energy emission from the jet, and source geometry in general, another question is regarding the disc inclination and whether the inner disc is aligned with the binary orbit, if it is set by the BH spin, or if the inner disc precesses, which has been suggested as a QPO mechanism \citep{ingram17} (see also \S\ref{sub:qpoorigin}).  Overall, there are still many open questions related to geometry of accretion in hard state, and answering the open questions related to the hard state accretion and source geometry is likely to require a combination of observations, theory, and numerical simulations.

\subsection{Corona origin / jet connection}
\label{sub:coronajet}

In Section~\ref{sec:overview}, we described several key observations that demonstrate a connection between the hot, optically thin corona at $R<100 R_{\rm g}$ and the relativistic jet observed at much larger distances ($10^{7-8} R_{\rm g}$). These include:
\begin{itemize}
    \item The X-ray/radio flux correlation in Active Galactic Nuclei and XRBs, known as the `fundamental plane' (\cite{merloni03}; \S\ref{sub:radionirjet})
    \item The low reflection fraction observed in the hard state, which has been interpreted as due to the corona/jet base accelerating away from the accretion disk. This beams much of the X-ray emission away from the disk (\cite{markoff05}; \S\ref{sub:broadironlines}).
    \item The close temporal coincidence of the Type~B QPOs during the state transition, and radio flares from the jet (\cite{Homan2020}; \S\ref{sub:qpo}).
    \item The correlation between the size of the X-ray corona (probed by X-ray reverberation), and the radio luminosity (\cite{wang21,demarco21}; \S\ref{sub:lags}).
\end{itemize}

Because of these and other observables, several models have been proposed that relate the corona and the jet. For instance, \cite{markoff05} suggest that the X-ray corona {\em is} the base of the relativistic jet. Thermal seed photons (and possibly also synchrotron photons from electrons in the jet) will be inverse-Compton upscattered in the jet base to produce the observed hard X-ray photons. This model maintains a spatially compact corona/jet-base that is required to explain the broad reflection features. \cite{kylafis08} and \cite{reig18} suggest a similar model where disk photons are upscattered in a jet, though require that the low-frequency hard lags (Section 3.1.6) are due to light travel time delays in the jet, and therefore require a much larger, extended jet. \cite{petrucci08} also suggest a connection between the hard X-ray emitting region and the relativistic jet, but propose that the hard X-rays originate in the plane of the accretion disk, in the so-called jet-emitting disk (JED), a highly magnetized, low-density inner accretion flow. The large large-scale jets then tap the mechanical energy released in the JED, via the Blandford-Payne mechanism \cite{blandford82}. The observational implication of the JED model is a truncated thin disk in the hard state, which is distinct from the jet-base model where the thin disk can extend to the ISCO. Because of these different observational predictions, future mutli-wavelength spectral-timing observations, paired with improved numerical simulations will reveal how the corona and jet are coupled.

\subsection{Origin of gamma-ray tail} 
\label{sub:gammarayorigin}

The most popular models that have been put forward to explain the origin of the high-energy tail in the hard and hard intermediate states of BHBs are Comptonization from a non-thermal electron population and synchrotron emission from a compact jet. The energy dependent polarization measurement for Cyg~X$-$1 above 400 keV nearly reaches the theoretical maximum for synchrotron radiation with ordered magnetic fields and strongly favors a jet origin beyond this energy as neither thermal nor non-thermal Compton scattering would provide polarization fractions close to the observed 75\%.  A jet origin for Cyg~X$-$1's MeV emission has interesting physical implications as it implies magnetic field strengths above the equipartition and a very efficient particle acceleration \cite{Zdziarski2014}.  There are no other reports of polarization measurement in hard X-rays/soft \gray s for any other BH sources, and whether the high polarization is a peculiarity of Cyg~X$-$1, or a common phenomena in all BH sources  remains to be seen.  A future MeV polarimeter such as the future {\em Compton Spectrometer and Imager (COSI)} \citep{tomsick21} will allow for sensitive polarization measurements of BH transients.

The case for Compton scattering from a non-thermal population of electrons have already been put forward to explain the tail in the soft state, and also have been invoked in the hard state as well. For lower quality data (e.g. during the outburst decays with lower luminosity) the presence of non-thermal or hybrid Comptonization is implied by poor fits with thermal only Comptonization models, and a significant improvement in reduced $\chi^{2}$ with addition of a non-Maxwellian electron energy distribution (using the $eqpair$ model \cite{Coppi1999}, for example).  With the much higher quality data, such as the case for \maxis, the picture becomes more complicated. The analysis of SPI data at different energy bands indicated a two component emission, modeled as a thermal Comptonization plus a cut-off power-law with a claim that the power-law has possibly a jet origin as it is only prevalent at high energies (100 - 300 keV) in the hard state. On the other hand, an analysis including lower energies using $NuSTAR$ and higher energies using SPI and PicSIT on $INTEGRAL$ led to an alternative picture with two Comptonizing regions, one closer to the BH which is producing both thermal and non-thermal Comptonization, and the other possibly over the disk at a larger radii with thermal Comptonization, without the need of a separate power-law component \cite{Zdziarski2021b}. 

Non-thermal tails in the electron energy distribution invoked to explain the hard tails would also invoke enhanced synchrotron radiation by the same electrons, possibly not only altering the seed photon energy distribution (much less than the disk photon energies in the hard state), but may also be producing near-infrared emission at the same energies where the jets could also contribute \cite{Veledina2013}.

\subsection{Origin of QPOs}
\label{sub:qpoorigin}

Many models have been forward to explain the origin of QPOs (for both HFQPOs and LFQPOs, even extending same models to explain the QPOs in neutron star systems), but in general, the models can be classifies as ``geometrical", or ``intrinsic". In the geometrical models, one of the accretion/ejection component's shape varies, and in the intrinsic models, the geometry is stable within the QPO oscillation timescale, however, the disk or the corona emission changes due to changes in the local properties. An extensive list of both geometrical and intrinsic models and their principles have been given in a recent review, \cite{Ingram2019}), and here only models relevant to general evolution of X-ray spectral and timing will be discussed.

The Type~C QPOs are observed almost every state, and currently the leading model to explain the origin of this QPO is the Lense-Thirring Precession (LTP) model arising from the frame dragging in the vicinity of a spinning BH \cite{Ingram2009}. This model assumes a thick accretion flow (or a corona) inside a  truncated disk misaligned with the equatorial plane of the BH causing a nodal precession of the entire inner flow due to the inequality between the orbital and vertical epicyclic frequencies. The epicyclic frequencies are the natural frequencies of a test mass disturbed slightly from the black hole equatorial plane, and in Kerr metric the orbital, radial and vertical frequencies are slightly different \citep{Ingram2019}. There are strong observational evidences for geometrical origin through the inclination dependence on the QPO RMS, and the iron line centroid frequency dependence on the QPO phase, both consistent with the predictions of LTP (though the latter is observed only in one source). There are also supporting arguments in favor of LTP, and against some of the intrinsic models. First of all, there is no sign of direct QPO modulation in the emission from the disk through frequency-resolved spectroscopy \cite{Sobolewska2006}, in contrast, the Type~C QPOs can be observed beyond 200 keV \cite{Ma2021}, therefore the modulated emission is coming from the corona. The LTP predicts a highest QPO frequency based on the spin, therefore this highest frequency should vary from source to source which is the case so far. It is important to stress that if LTP is indeed the correct model for Type~C QPOs, most BH disk axis should be misaligned with the spin axis, and should stay misaligned (with at least 5$^{\circ}$) for a long time as the accretion brings them back to alignment. There are observational and theoretical arguments in favor of misalignment which can be found in Ingram \& Motta 2019 \cite{Ingram2019} and references therein.  

Type~B QPO properties also depend on inclination, pointing to a geometrical origin, yet their RMS is higher for the lower inclinations systems whereas the opposite is observed for Type~C QPOs \cite{Motta2015}. One model associates Type~B QPO to the precession of the jet \cite{Motta2015}, as opposed to the precession of the thick flow in Type~C QPOs. 

For the HFQPOs, the 2:3 ratio in GRO~J1655$-$40 (and weaker detection claims in other sources) suggested models invoking resonances related to test particle orbits, and for the case of GRO~J1655$-$40, with the accompanying low frequency QPOs, a simple relativistic precession model has been suggested in which the LFQPO fundamental frequency, the lower, and the higher HFQPOs are the Lense-Thirring precession, periastron precession, and the orbital frequency at a characteristic radius, respectively \cite{Motta14}. Knowing all three frequencies allowed to determine the spin, characteristic radius and mass of the black hole, and the inferred mass is in agreement with the mass determined through optical observations. The HFQPO pair can also be explained within the framework of a precessing torus with frequencies corresponding to the breathing mode and vertical epicyclic frequency of the inner flow \cite{Fragile2016}.

\section{Future of black hole research in X-ray and Gamma-ray domain}
\label{sec:conclusion}

Our understanding of the accretion and ejection mechanisms around BHBs increased substantially in the last 25 years, with major contributions from \rxte\ followed by $Chandra$, \xmm, $INTEGRAL$ in X-rays and \gray s, accompanied with  multiwavelength observing campaigns to link the accretion processes to the launch of relativistic jets. In the last 10 years, we had another huge leap with \nicer, \nustar, and \hxmt, resulting a deeper understanding of X-ray spectral components and their relation to the timing properties.

We appear to be approaching yet another milestone to put the controversial issues discussed in \S\ref{sec:interpretation} to rest with the upcoming approved and concept missions,  and perhaps opening up new and exciting discoveries to be solved in the future. Specifically, the {\em Imaging X-ray Polarimetry Explorer} ($IXPE$, \cite{Weisskopf2016}) launched at the end of 2021 and enhanced {\em X-ray Timing and Polarimetry} ($eXTP$, \cite{Zhang2019}) to be launched in late 2027 will allow polarization measurements in X-rays, providing an opportunity to test specific predictions of QPO models and accretion geometry in the hard state \cite{DeRosa2019}. Even measuring stochastic polarization variability in short time scales may be possible in the near future with $IXPE$ (together with a large area instrument, such as $STROBE-X$) or directly with the $eXTP$ utilizing Polarimetry Focusing Array with the Large Area Detector \citep{Ingram17_2}. In addition, {\em COSI} (to be launched in 2025, \citep{tomsick21}) will measure polarization at soft \gray\ energies, to determine if the MeV and X-ray emission has a different origin as has been suggested by $INTEGRAL$ measurements of Cyg~X$-$1.

In addition to polarimetry, studies of highly ionized outflows via absorption line spectroscopy will be revolutionized by the launch of $XRISM$, a high-resolution X-ray microcalorimeter, set to launch in 2023 \citep{xrism}. $XRISM$'s energy resolution at hard X-rays will be unparalleled until the launch of $Athena$ in 2034. High-resolution spectroscopy is also expected to clarify the multi-component line profiles often seen as part of reflection components.

$eXTP$, and $STROBE-X$ (\cite{Ray2019} if recommended for implementation) would provide plenty of photons thanks to an order of magnitude increase in their effective areas compared to \rxte\ to study physics of accretion in strong gravity in unprecedented detail. The other advantages of these new generation large observatories over \rxte\ will be their low energy coverage extending well below that of \rxte, and better energy resolution allowing relating variability in absorption and emission lines to changes in fast variability properties without a need for coordinated observations with other satellites. These two missions offer complimentary characteristics to study BHBs, the $eXTP$ with capability allowing simultaneous polarimetric measurements with great timing and spectral coverage and $STROBE-X$ with excellent coverage in less than 2 keV with the X-ray Concentrator Array with an area an order of magnitude larger than \nicer. Both instruments will carry a very large area, very large field of view Wide Field Monitor (WFM) \citep{Hernanz18} allowing not only detecting new transient sources when they are still dim, but also allowing following state transitions in detail with the WFM data only. 

In addition to future advancements in X-ray instrumentation, the last decade has shown the importance of coordinated multi-wavelength campaigns, especially pairing with fast photometric instruments in the optical, IR and sub-mm. While often challenging to coordinate, these campaigns help us probe the causal connection between the accretion disk, corona and relativistic jet. Finally, just as we improve telescope technology, we continue to see fast improvement in numerical simulations of accretion flows in X-ray binaries, including the effects of general relativity and radiation in the magneto-hydrodynamics. As GPU-accelerated simulations continue and improve in the future (e.g. \cite{liska,Dexter2021}), we will better be able to connect disk theory to observations.

\section{Cross-References}

Volume 1: History of X-ray Astrophysics, Astrosat, ATHENA, Chandra, eXTP, Insight-HXMT, MAXI, NICER, Swift, IXPE The Imaging X-ray Polarimetry Explorer, XMM-Newton, Study Concepts for future mission in High Energy Astrophysics.

Volume 2: History of Gamma-ray Astrophysics, The COMPTEL instrument on the CGRO mission, The INTErnational Gamma-Ray Astrophysics Laboratory (INTEGRAL), Low mass X-ray binaries, High mass X-ray binaries, Overall accretion disk theory, Black holes: Accretion processes, Fundamental physics with black holes, Probing black-hole accretion through time variability, Tests of General Relativity using black hole X-ray data, X-ray polarimetry-timing, Soft gamma-ray polarimetry with COSI using maximum likelihood analysis.

\section{Acknowledgements}

The authors thank Jingyi Wang for providing data and making plots of MAXI~J1820+070 found in Figures \ref{fig:maxi_hid}, \ref{fig:maxi_psd} and \ref{fig:maxi_lag}.  The authors thank Dr. Douglas Buisson for providing MAXI~J1820+070 energy spectra, which were used in the making of Figures \ref{fig:maxi_spe} and \ref{fig:cartoon}.  The authors thanks Dr. Javier Garcia for providing a first version of Fig.~\ref{fig:cartoon}.

JAT acknowledges partial support from NASA under grant 80NSSC20K0644. EAK acknowledges partial support from NASA~ADAP grant 80NSSC17K0515.


%
%
%

\end{document}